# Pressure-Induced Changes in Structure, Magnetic Order and Development of Superconductivity in the Ferromagnetic Topological Insulator MnBi$_8$Te$_{13}$

S. Huyan[1,2*†], T. Qian[3*], L. Wang[1,2], W. Bi[4], F. Xue[5], D. Zhang[6], C. Hu[3], B. Kalkan[7,8], Y. Huang[9], Z. Li[1,2], A. Das[1,2], J. Schmidt[1,2], R. A. Ribeiro[1,2], T. J. Slade[1,2], N. Ni[3†], P. C. Canfield[1,2†], and S. L. Bud'ko[1,2†]

[1] *Ames National Laboratory, US DOE, Iowa State University, Ames, Iowa 50011, USA*
[2] *Department of Physics and Astronomy, Iowa State University, Ames, Iowa 50011, USA*
[3] *Department of Physics and Astronomy and California NanoSystems Institute, University of California, Los Angeles, California 90095, USA*
[4] *SmartState Center for Experimental Nanoscale Physics, Department of Physics and Astronomy, University of South Carolina, Columbia, SC 29208, USA*
[5] *Department of Physics, University of Alabama at Birmingham, Birmingham, AL 35294, USA*
[6] *Center for Advanced Radiation Sources, The University of Chicago, Chicago, Illinois 60637, USA*
[7] *Department of Earth and Planetary Sciences, University of California, Santa Cruz, CA 95064, USA*
[8] *Advanced Light Source, Lawrence Berkeley National Laboratory, Berkeley, CA, USA*
[9] *Computational Quantum Matter Research Team, RIKEN Center for Emergent Matter Science (CEMS), Wako, Saitama 351-0198, Japan*

† shuyan@iastate.edu, nini@physics.ucla.edu, canfield@ameslab.gov, budko@ameslab.gov.

Abstract:

We report a comprehensive study of pressure-induced evolution of the magnetism and development of superconductivity (SC) in MnBi$_8$Te$_{13}$, a promising ambient pressure, ferromagnetic (FM) topological insulator candidate. By employing high-pressure electrical transport, magnetoresistance, DC magnetic susceptibility, and X-ray diffraction measurements, we construct a detailed temperature-pressure phase diagram. At ambient pressure, MnBi$_8$Te$_{13}$ exhibits FM ordering with an easy-axis along the *c*-axis which is progressively suppressed under pressure and replaced by an antiferromagnetic (AFM) order. Density functional theory calculations predicted an evolution from FM to a G-type AFMg2 phase near 5 GPa. Above 16.6 GPa, a bulk SC state emerges with a maximum transition temperature ~6.8 K, as confirmed by resistance and magnetic susceptibility measurements. This pressure-induced SC may co-exist with another AFM dome that shows a weak anomaly in the transport data. In contrast, our work on MnBi$_6$Te$_{10}$ shows no SC up to 40 GPa. Indeed, in contrast to MnBi$_8$Te$_{13}$, the MnBi$_2$Te$_4$(Bi$_2$Te$_3$)$_n$ compounds with n < 3, didn't exhibit SC, highlighting the crucial role of Mn concentration in stabilizing SC. The observation of pressure-induced FM-AFM-SC transitions in MnBi$_8$Te$_{13}$ not only establishes it as a rare Mn-based SC but also provides a platform to study the interplay between magnetism, SC, and potentially nontrivial band topology in correlated magnetic materials.

**Introduction:**

Magnetic topological insulators (MTIs) represent a distinct class of materials that uniquely blend magnetic and topological properties [1-6]. Among these, MnBi$_2$Te$_4$ was the first material predicted to be an intrinsic MTI [7], leading to increased attention to the Mn-Bi-Te series due to its potential applications in quantum computing and spintronics [8]. Within the Mn-Bi-Te series, manganese (Mn) and bismuth (Bi) atoms are arranged in layered structures, giving rise to various novel magnetic phenomena. These include A-type antiferromagnetism (AFM) in MnBi$_2$Te$_4$ [9], the promising axion insulator candidates MnBi$_4$Te$_7$ [10], and MnBi$_6$Te$_{10}$ [11], along with the notable MnBi$_8$Te$_{13}$, which is one of the topological insulator candidate exhibiting ferromagnetic (FM) ordering [12]. The diverse range of compounds with different stacking sequences in the Mn-Bi-Te system provides a fertile platform for exploring the interplay between interlayer interactions, topological properties, magnetic states, and other novel ground states.

In light of these intriguing magnetic properties, researchers are exploring ways to induce superconductivity (SC) in intrinsic MTIs, as the coexistence of magnetism, topology, and SC could unlock exotic quantum states [13], enable Majorana fermions for quantum computing [14], and provide insights into potential unconventional SC mechanisms [15]. One common approach involves introducing electron or hole carriers through doping [16,17]. By selectively introducing additional charge carriers into the material, the Fermi level can be tuned, leading to a transition from an insulating to a SC state. Additionally, proximity-induced SC can be achieved by placing the MTI in close proximity to a SC material, allowing the SC order parameter to penetrate the MTI and induce SC in its surface states. Notably, in the nonmagnetic 124 compounds within the same structural family of MnBi$_2$Te$_4$, bulk SC was reported in In-doped SnBi$_2$Te$_4$ crystals [18] and PbBi$_2$Te$_4$ [19] crystals (maximum $T_c$ ~ 1.85 K and ~ 2.06 K, respectively) as well as in pure PbBi$_2$Te$_4$ under high pressure [20]. Conversely, no SC has been observed in Mn-based 124 compounds, as Mn tends to form compounds that exhibit localized magnetic moments, which are typically unfavorable for SC. For example, MnB$_2$Te$_4$ and MnBi$_4$Te$_7$ do not display SC despite attempts at chemical doping [21,22] or high pressure [23], even when magnetic ordering is fully suppressed—an observation that contrasts with unconventional SC such as cuprates [24], iron-pnictides [25-26], and heavy-fermion SCs [27], where suppressing magnetism often leads to SC. This suggests that in Mn-based 124 compounds, strong electron correlations and localized magnetic moments play a fundamentally different role, making the emergence of SC less favorable. Although SC is generally rare in Mn-based materials, notable exceptions such as MnP [28] and MnSe [29], demonstrate that external pressure can stabilize SC, underscoring the subtle and material-specific balance between magnetism and pairing in these systems. This situation closely parallels to that of the local-moment rare-earth intermetallics, where SC is typically absent in Dy-, Ho-, Er-, or Tm-based compounds, yet remarkable exceptions exist. For example, the $RRh_4B_4$ [30] and $RNi_2B_2C$ [31,32] families reveal that SC can not only coexist with local-moment magnetism, both paramagnetic and AFM, but in some cases remain robustly conventional, as in the BCS electron-phonon SC of the $RNi_2B_2C$ system

Whereas detailed $T$-$p$ phase diagrams for $MnBi_2Te_4$ and $MnBi_4Te_7$ have been reported [23], no such studies have been conducted for $MnBi_6Te_{10}$ and $MnBi_8Te_{13}$. To fill this gap, we measured the temperature-dependent resistance of $MnBi_6Te_{10}$ and room temperature powder x-ray diffraction under pressures up to 40 GPa and 23.5 GPa, respectively. These compounds feature a lower Mn concentration compared to previously studied $MnBi_2Te_4$ and $MnBi_4Te_7$, providing an opportunity to explore how reduced Mn content influences magnetic and SC properties under pressure. Whereas $MnBi_6Te_{10}$ behaves similarly to previous reports on $MnBi_4Te_7$ [23], $MnBi_8Te_{13}$ exhibits a suppression of FM order, transition into AFM states, and ultimately the onset of SC, possibly co-existing with some form of AFM ordering, under high pressure. The density functional theory (DFT) calculations indicate that the AFM order above ~ 5 GPa likely adopts an $AFM_{g2}$ configuration - a variant of G-type AFM ordering. And remarkably, among the Mn-Bi-Te ($MnBi_2Te_4(Bi_2Te_3)_n$, $n \leq 3$) family, only $MnBi_8Te_{13}$ displays SC under pressure, emphasizing the important role of Mn dilution and the corresponding increased Mn-Mn distance in weakening magnetic interactions and enabling SC pairing. This trend underscores how variations in Mn content directly impact electronic correlation strength and magnetic exchange pathways. Our findings position $MnBi_8Te_{13}$ as a promising platform for investigating SC in topological systems with strong correlations.

**Results**

Figure 1 presents a concise summary of our primary experimental results. After reviewing these results, we will present and discuss the specific data sets used to create the *T-p* phase diagrams shown. Fig. 1(a) plots (i) magnetic and superconducting (SC) transition temperatures (ii) structures, as a function of applied pressure on MnBi$_8$Te$_{13}$. The compound exhibits unusual, complex FM-AFM-SC transitions within the structural evolution of rhombohedral to monoclinic to cubic with increasing pressure. In the pressure region of the rhombohedral structure (P$_I$), the sample undergoes a discontinuous transition from the FM state to another magnetic phase that, as will be discussed below, closely resembles an antiferromagnetic (AFM) state, AFM1, above ~ 4 GPa. As the sample approaches and undergoes structural transition from the rhombohedral to the monoclinic phase (P$_{II}$), a second AFM dome (AFM2) could be observed with $T_N$ slightly higher than AFM1. Upon applying pressure above ~16 GPa, the onset of the structural transition from monoclinic (P$_{II}$) to cubic (P$_{III}$) phase, the SC with maximum $T_c$ ~ 6.8 K was observed accompanied by the monoclinic to cubic (P$_{III}$) transition.

We also studied the electrical transport properties on MnBi$_6$Te$_{10}$ under pressure which is a much sparser data set than the one we assembled for the 1-8-13, as such only allows us to make qualitative statements. Our results reveal that MnBi$_6$Te$_{10}$ undergoes a gradual suppression of magnetic order under pressure. Starting from an AFM ground state at ambient pressure with a transition temperature of around 10 K, the ordering temperature decreases monotonically to approximately ~3.7 K at ~8.8 GPa, as shown in Fig. 1(b), S8 and S9. Although the signature of the magnetic transition weakens as pressure is increased to 8.8 GPa, we can detect it over the whole P1 pressure range (Fig. 1b). It is also worth noting that whereas we did not perform a detailed investigation of the magnetic structures under pressure, previous studies suggest that MnBi$_6$Te$_{10}$ undergoes a pressure-induced transition from A-type AFM to a quasi-two-dimensional ferromagnetic (FM) state with a reduced saturation field above 1 GPa [33]. Nevertheless, a key result from our high-pressure transport measurements is that no sign of SC is detected down to 1.8 K for pressures up to 30 GPa.

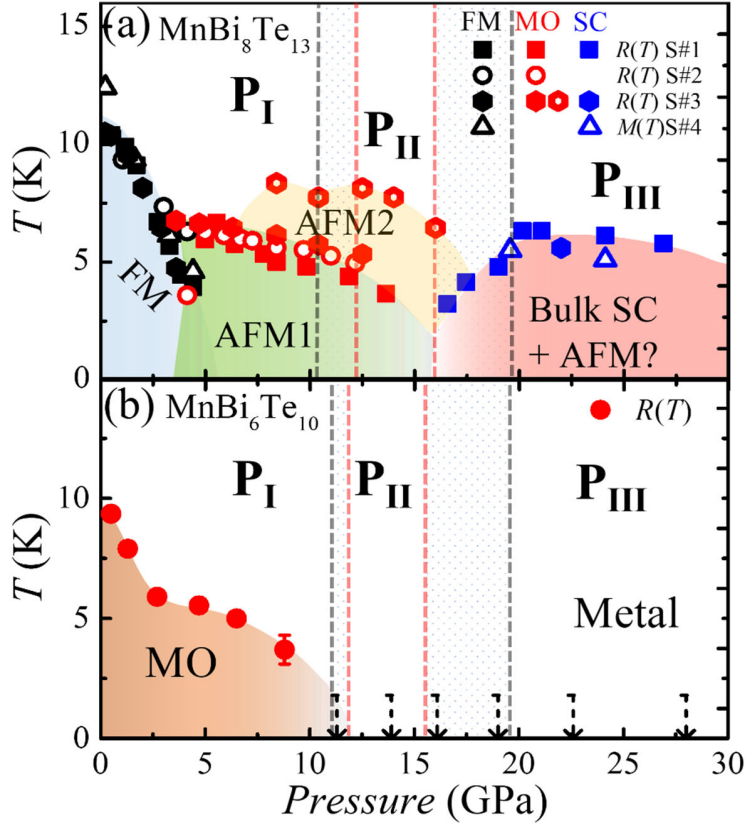

Fig. 1 Temperature - pressure ($T$ - $p$) phase diagram of (a) $MnBi_8Te_{13}$, and (b) $MnBi_6Te_{10}$. The black and red dashed lines divide the diagrams into three regions—PI, PII, and PIII—corresponding to distinct crystallographic phases. The pressure intervals between these dashed lines represent mixed-phase regions (PI+PII and PII+PIII), as identified by powder X-ray diffraction measurements. For $MnBi_8Te_{13}$, samples S#1 to S#4 are cleaved pieces from the same batch. S#1–S#3 were used for resistance measurements, and S#4 for magnetization measurements. Given that we cannot detect AFM features in our $M(T)$ data and given that the SC feature in $R(T)$ obscures any possible AFM feature with $T_N < T_c$, it is possible that there is AFM ordering at or below $T_c$ for P > 17 GPa in $MnBi_8Te_{13}$. All symbols in (a) indicate magnetic or superconducting transition temperatures obtained from three independent $R(T)$ measurements and one $M(T)$ measurement across four different samples. In the case of $MnBi_6Te_{10}$, one sample was used for $R(T)$ measurements up to 40 GPa. Features associated with the suppression of the ambient pressure AFM transition are shown in red solid circle. At pressures above 8.8 GPa, we do not detect any features of transitions in the $R(T)$ data and instead just show, in black, stars consistent with our base temperature.

Figure 2 presents the temperature-dependent electrical resistance and magnetization of $MnBi_8Te_{13}$. The sample exhibits metallic behavior in the whole pressure range. Increasing the pressure induces a continuous suppression of the overall magnitude of temperature-dependent resistance $R(T)$ in the first pressure region from 1.2 GPa to 4.4 GPa, during which the FM transition temperature $T_C$ monotonically decreases with pressure. A resistance drop feature is observed below $T_C$ due to the loss of spin disorder scattering, and its magnitude appears to evolve with pressure (Fig. 2(b)). Correspondingly, magnetization in the ordered state monotonically decreases with pressure, as shown in Fig. 2(e). A clear deviation between the Zero-Field-Cooled (ZFC) and Field-Cooled (FC) curves emerges at lower temperatures below $T_C$. This behavior has been reported previously and has been attributed to the formation of magnetic domains at low temperatures [12].

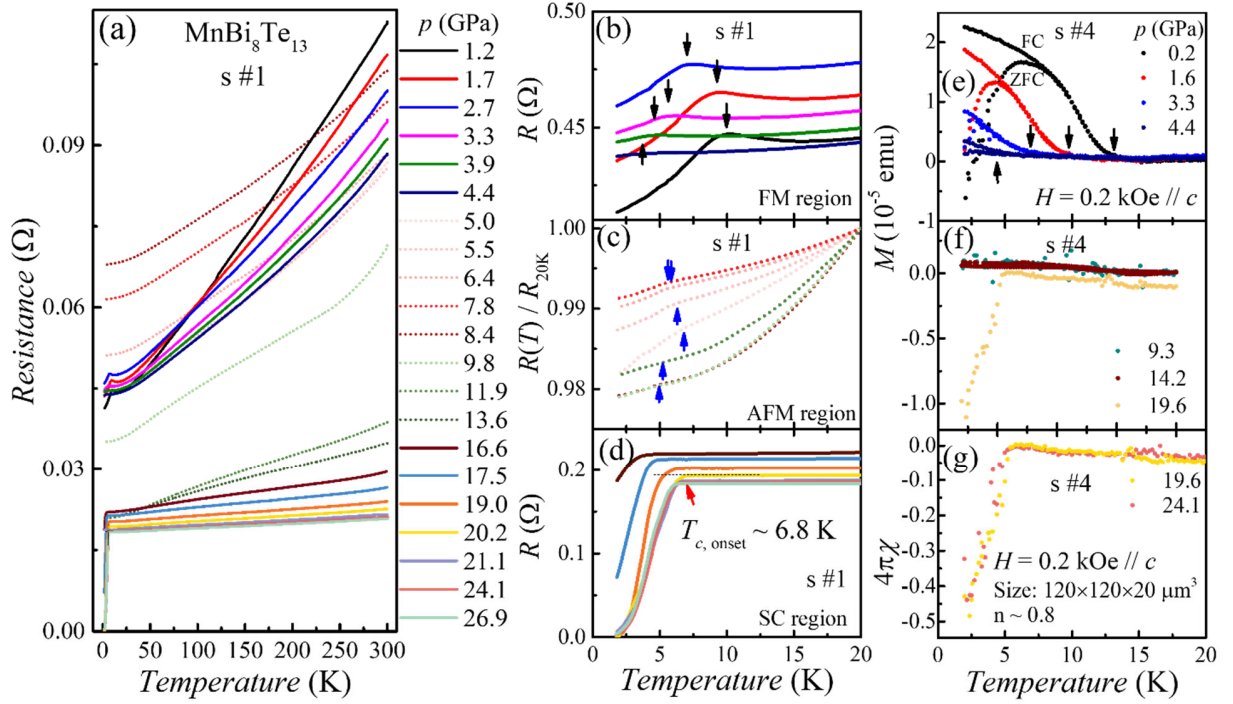

Fig. 2 (a) Temperature-dependent resistance as function of pressures of sample 1 (s#1) of MnBi$_8$Te$_{13}$. (b), (c), (d) are enlarged plots of different pressure regions of s#1 marked as FM, AFM and SC regions, respectively. (e), (f) and (g) are the magnetic susceptibilities at corresponding regions on s#4.

In the pressure range of ~4 to ~8 GPa, as shown in Fig. 2(a, c and f), the overall resistance increases with pressure, and the $R(T)$ curves become progressively flatter, suggesting a suppression of temperature-dependent scattering mechanisms and a potential reconstruction of the electronic structure. In addition, a transition-like feature is observed at ~6 K, where a slight resistance drop appears in sample 1 (s#1, Fig. 2(c)), whereas a clear upturn is seen at the same onset temperature in s#2 and s#3 (see Fig. 3 (b), (c) and Fig. S1-S7 in SI). This behavior is reminiscent of the formation of a magnetic superzone gap due to Fermi surface nesting [34-38], a phenomenon typically associated with modulated or incommensurate AFM states rather than the simple commensurate A-type AFM at ambient pressure observed in MnBi$_2$Te$_4$, MnBi$_4$Te$_7$, and MnBi$_6$Te$_{10}$ [39,40]. From ~8 GPa to ~ 14 GPa, a significant drop in the overall resistance is observed, closely linked to the structural transition from the rhombohedral (R-3m) to a monoclinic phase (indexed to A2/m, shown in Fig.1 (a), Fig. 4(a), and Fig. S10 in SI). Whereas no distinct multiple transitions are observed in s#1 due to the weak magnetic transition signatures, s#2 exhibits a clear two-step transition, starting at 8.4 GPa (Fig. 3(c)), persisting up to 10.4 GPa, before fully transitioning to the higher $T_N$ state with weak transition features at 12.5 GPa. The transition feature is gradually suppressed with increasing pressure up to 16 GPa. In the entire AFM-like pressure region (see Fig. 2(f)), the magnetic susceptibility signal falls below the detection limit of our measurement system (~$10^{-6}$ emu). Whereas this precludes direct observation of the magnetic transition, it could support the scenario of an AFM ground state in this regime.

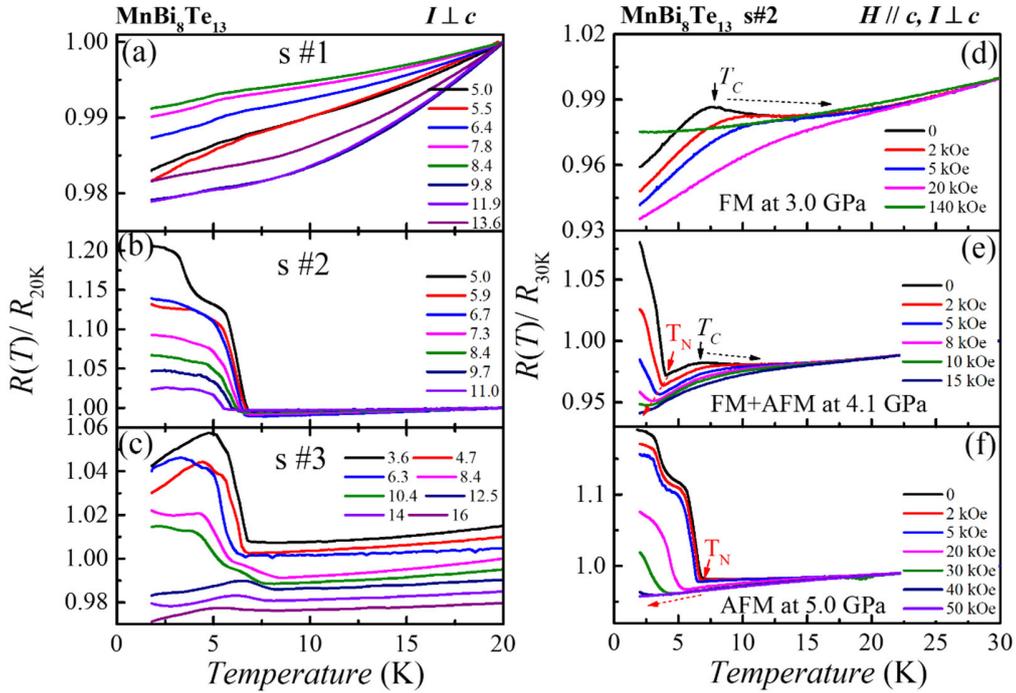

Fig. 3 Temperature-dependent normalized resistance at various pressures in possible antiferromagnetic regions. (a), (b), and (c) are for samples s#1, #2, and #3, respectively. The data is normalized at 20 K. the curves in (c) are shifted so as to better show the evolution of magnetic transition with the pressure. (d)-(f) temperature dependence of the resistance at various magnetic fields, at 3.0 and 4.1 GPa, and 5.0 GPa, respectively, along *c* direction. The current is parallel to *ab* plane.

Remarkably, a distinct resistance drop emerges at low temperatures at 16.6 GPa, as shown in Fig. 2(d), suggesting the onset of a non-bulk (since temperature dependent resistance does not reach a zero value) SC state. With increasing pressure, a robust SC phase develops between 19 GPa and 26.9 GPa, reaching a maximum $T_c$~6.8 K with near-zero resistance (Fig. 2(d)). The broad transition width may stem from pressure inhomogeneity or the material's intrinsic tendency to exfoliate, leading to possible variations in the local stress distribution. To further probe the SC nature, we performed magnetic susceptibility measurements at 19.6 GPa and 24.1 GPa, under an applied field of 0.2 kOe (Fig. 2(g)). The results reveal a clear unsaturated diamagnetic transition below $T_c$ persisting down to 2 K. The estimated shielding volume fraction at 2 K remains around 50 % at both pressures (with an estimated demagnetization factor $n = 0.8$). Given that the diamagnetic transition does not saturate, it is likely that the shielding fraction continues to increase at lower temperatures, reinforcing the notion that the pressure-induced SC in $MnBi_8Te_{13}$ is on the verge of achieving a true bulk SC state.

Since our data cannot resolve any other feature below the SC $T_c$, we cannot accurately comment on the possibility of a coexistence of AFM and SC. The extremely weak magnetic signals in the intermediate pressure range, combined with the onset of resistive transitions and diamagnetic responses, leave open the scenario where SC may coexist with an AFM-ordered background as shown in Fig. 1(a). High-pressure neutron diffraction or µSR studies will be crucial to resolve this question. Given that Mn very rarely sheds its moment in intermetallic compounds, the odds are that there is still Mn magnetism (disordered or AFM) in the high pressure, SC state.

High-pressure synchrotron X-ray diffraction measurements on $MnBi_8Te_{13}$ and $MnBi_6Te_{10}$ reveal a two-step structural transition sequence under compression. As indicated in Fig 4 and Fig. S10 the first transition, occurring near 10 GPa, involves a change from the ambient-pressure rhombohedral *R-3m* structure to a low-symmetry monoclinic phase, possibly to *A2/m*. A second transition emerges at ~16 GPa, suggesting a transformation from the monoclinic phase to a high-symmetry cubic (*Im3m*) structure. Remarkably, the structure of $MnBi_8Te_{13}$ is fully reversible upon pressure release, confirming its crystalline-not amorphous-nature.

These structural features are qualitatively consistent with those previously reported for $MnBi_4Te_7$ by Pei et al. [23], which similarly exhibits a rhombohedral to monoclinic to cubic evolution under pressure. In contrast, $MnBi_2Te_4$, whereas also undergoing the initial rhombohedral to monoclinic transition, fails to form a high-pressure cubic phase. Instead, it becomes amorphous above ~16 GPa, as evidenced by the broadening and the disappearance of Bragg peaks. This amorphization under pressure is not reversible. This difference suggests that higher-n Mn-Bi-Te members (n = 2, 3, 4) possess enhanced structural resilience under pressure, likely due to the increased number of $Bi_2Te_3$ spacer layers, which help suppress amorphization and stabilize crystalline transitions.

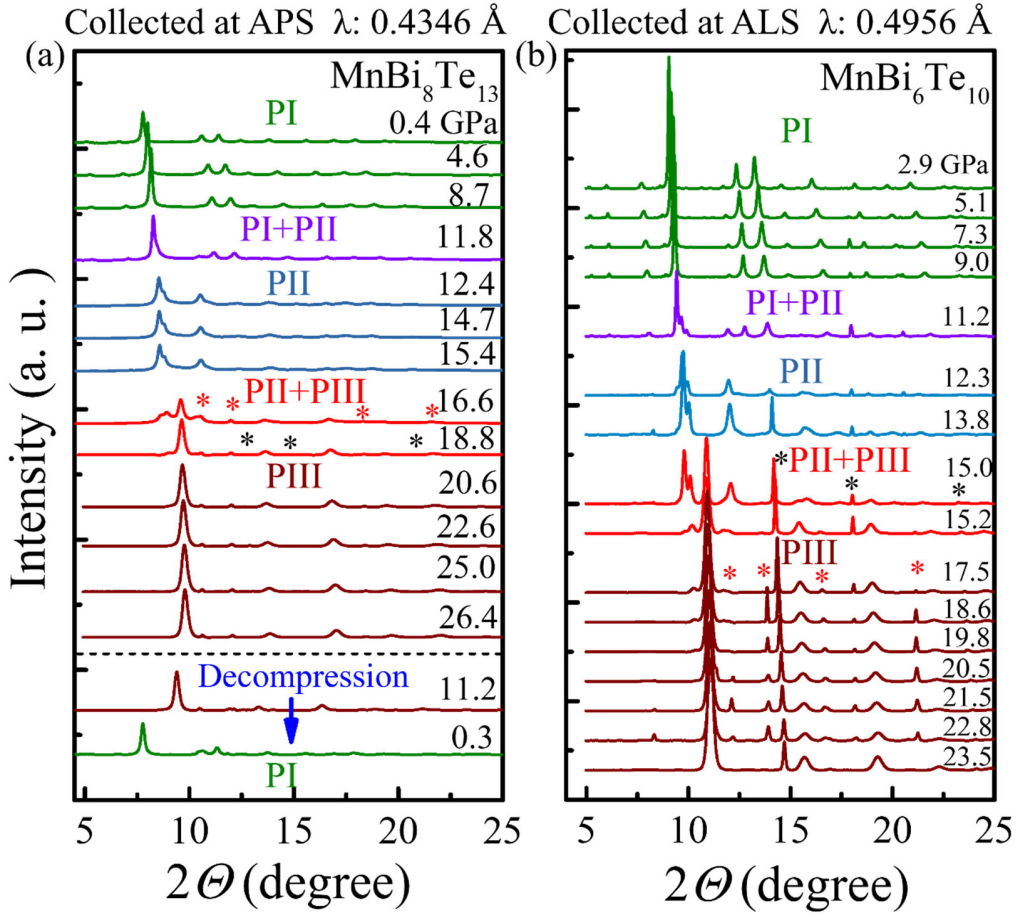

Fig 4 Selective powder x-ray diffraction data under various pressures on (a) $MnBi_8Te_{13}$, and (b) $MnBi_6Te_{10}$. The red and black stars in the figure represent the diffraction peaks of rhenium (from gasket), and crystalized neon (pressure transmitting medium), respectively. As discussed in the Experimental Methods section, the two X-ray diffraction measurements—performed at APS and ALS—used different X-ray wavelengths of 0.4346 Å and 0.4956 Å, respectively.

Interestingly, despite their distinct stacking sequences, $MnBi_4Te_7$, $MnBi_6Te_{10}$, and $MnBi_8Te_{13}$ all undergo structural transitions at nearly identical critical pressures (~10 GPa and ~16 GPa). This convergence points to a common underlying lattice instability in the high-n members of the Mn-Bi-Te series, although its microscopic origin remains to be clarified. Such a shared structural response to pressure provides a natural basis for comparing their low-temperature properties, particularly given the fact that only $MnBi_8Te_{13}$ exhibits pressure-induced SC.

**Evolution of the magnetic structure under pressure in $P_I$ phase**

Temperature-dependent resistance measurements at various external fields were performed on the MnBi$_8$Te$_{13}$ single crystal (s#2), with the current running along the *ab*-plane and the magnetic field direction parallel to the *c*-axis. At low pressures, e.g., 3.0 GPa shown in Fig. 3(d), when the magnetic field is increased, the PM-FM transition broadens and shifts to higher temperatures. This is consistent with the resistive signature of the FM transition in a metallic sample. The sample seems to have the FM ground state and high-pressure AFM state coexisting at 4.1 GPa. As shown in Fig. 3(e), both the resistance drop feature corresponding to the FM transition and the resistance upturn that corresponds to the new magnetic ordering could be observed. Upon the increase of the external field, the FM transition shifts to higher temperature and broadens, whereas the AFM transition is suppressed with the field which aligns with the spin polarization of the AFM ordering. Upon further increasing the pressure, when the sample enters the high-pressure magnetic ground state, e.g. at 5.0 GPa as shown in Fig. 3(f), the transition temperature $T_N$ behaves similar to the $T_N$ at 4.1 GPa, which indicates the possibly AFM state. It is noteworthy that at 5.0 GPa, a two-step resistance upturns could be also observed at zero field, both onset temperatures are suppressed with the magnetic field, indicating possible multiple incommensurate AFM states. The feature is faint at 5.9 GPa. Similar measurements were also performed on s#3 with external field along *ab*-plane and perpendicular to the direction of the current, where same results as s#2 were observed, as shown in Fig. S7.

Transverse MR measurements were performed under high pressure with the external field applied along the crystallographic *c* and *ab* directions, revealing a significant evolution of the magnetic structure from low to high pressures. At 0.5 GPa, see Fig. 5 (a), (b), moderate negative MR is observed for both field orientations, with a hysteresis of approximately 0.5 kOe. Full spin polarization is achieved above 10 kOe for magnetic fields along the *c*-axis and above ~20 kOe along the *ab*-plane, as indicated by the minima in the MR curves. This anisotropic field response supports the assignment of the *c*-axis as the magnetic easy-axis. Beyond these fields, a gradual upturn in MR is observed, which indicates a normal-state orbital MR contribution once spin disorder scattering is fully suppressed. The MR results at 0.5 GPa are qualitatively consistent with the single-crystal neutron diffraction which confirmed the long-range FM ordering below ~12 K along the *c*-axis at ambient pressure [12].

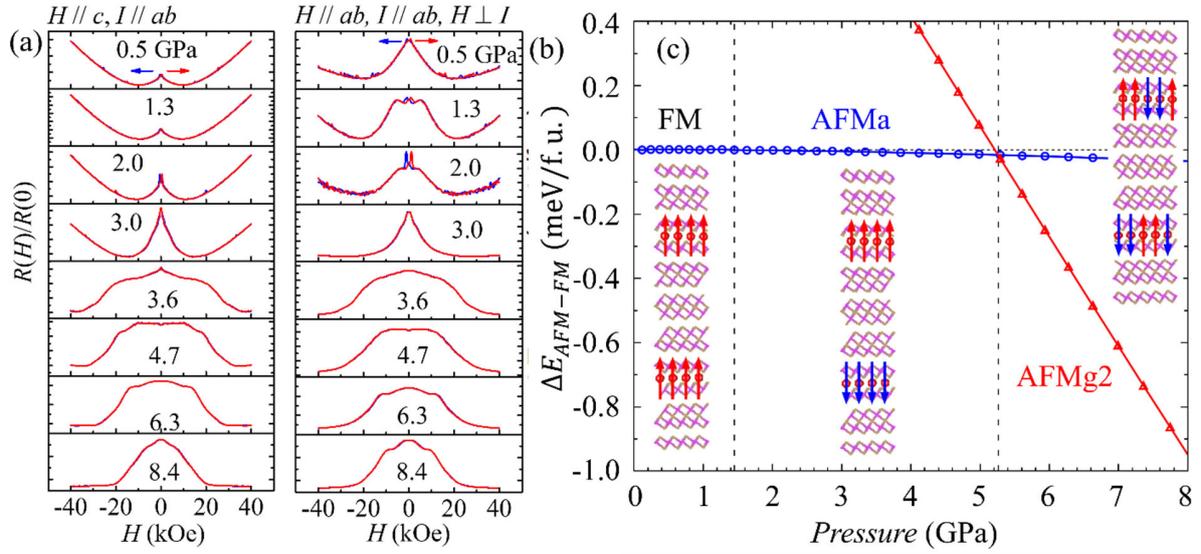

Fig. 5. (a) and (b) Anisotropic transverse magnetoresistance as functions of magnetic fields on s#3 along different directions, with current direction unchanged. Blue and red curves are the date measured with magnetic fields sweeping from 40 kOe to -40 kOe, and from -40 kOe to 40 kOe, respectively. (c) Three different intralayer magnetic configurations, including intralayer ferromagnetic ordering, A-type AFM (AFMa) ordering and AFMg2 ordering.

At 1.3 GPa, the MR curve for $H//c$ exhibits a behavior similar to that at 0.5 GPa, with a slight increase in the field required for full spin polarization. However, when the field is applied along the *ab*-plane, a clear cascade of features occurs below 10 kOe. This behavior may reflect spin canting or a field-induced reorientation from a *c*-axis-aligned FM configuration to a state with magnetic moments gradually rotating into the *ab*-plane, consistent with anisotropic FM or a crossover into a saturated paramagnetic regime.

Above 3.6 GPa, the *MR* curves for fields along both *c* and *ab* directions become nearly identical, suggesting an increasing isotropy in the spin structure under high pressure. The field required to induce the metamagnetic-like transition decreases monotonically with pressure, in agreement with the suppression of $T_N$, shown in the *T* - *p* phase diagram (see Fig. 1(a)). Furthermore, at and above 6.3 GPa, a two-stage transition feature emerges, which may reflect the presence of multiple metamagnetic transitions arising from different AFM configurations with comparable energies. This is plausible given that two distinct AFM ground states are already stabilized at zero field. High-pressure neutron diffraction studies would be essential to further resolve the nature of these closely competing magnetic phases. Although our transport and magnetization data are consistent with a FM ground state at low pressure, the presence of multiple pressure-induced AFM phases suggests that the FM region may also host finite-q spin textures or modulated order. Such possibilities cannot be simply excluded based on the present measurements. High-pressure neutron diffraction or µSR will be required to resolve the precise nature of the magnetic state.

The compounds in Mn-Bi-Te family are known for the delicate interlayer magnetic coupling between the stacked FM van der Waals layers. However, inside the same van der Waals layer with a 2D triangular Mn lattice, there are also competing magnetic interactions. Early DFT calculations [41,42] on MnBi$_2$Te$_4$ considering G-type AFM to preserve the non-symmorphic time-reversal symmetry for surface Dirac cone also found these G-type AFM are close in energy to the A-type AFM ground state. Indeed, the inelastic neutron scattering study [43] on MnBi$_2$Te$_4$ found a sizable next-nearest neighbor AFM coupling besides the nearest neighbor FM interaction to make the 2D triangular Mn lattice near frustration and a possible complex magnetic phase diagram. DFT calculation [41] and Monte Carlo simulation [43] showed that the intralayer double-row (or up-up-down-down stripe) are more stable than the single-row AFM (1.9 vs. 4.7 meV/f.u. in DFT). A later DFT study [44] on MnBi$_2$Te$_4$ found that pressure can make the single-row AFM become favorable to AFMa.

Here we study the pressure effect on the magnetic phase stability of MnBi$_8$Te$_{13}$ by considering the FM ground state, AFMa and AFMg2 with the double-row in the latter being more stable than the single-row. The calculation was performed within the pressure range where the structure remains in PI phase. The energies of different intralayer magnetic configurations of FM, A-type AFM (AFMa) and G-type AFM (AFMg2) magnetic configurations as function of pressure are shown in Fig. 5 (c). The results reveal that the energy difference between the AFMa and FM states is very small and decreases nearly linearly with pressure. This difference crosses zero around 1.3 GPa, potentially indicating that a pressure-driven transition from FM to a non-FM state with small energy differences between competing AFM configurations, suggesting the presence of frustrated magnetism or complex spin textures, which is qualitatively consistent with the MR results. It is noteworthy that the gaps between FM and A-type AFM are very small, probably indicating the high degeneracy of the different magnetic configurations. Such a scenario may underline the MR anomaly observed near 1.3 GPa under an in-plane magnetic field.

Interestingly, whereas the AFMg2 configuration is significantly less stable than FM at low pressure, the energy difference between AFMg2 and FM states decreases sharply with increasing pressure and AFMg2 becomes more stable above ~5 GPa. This trend aligns well with our MR measurements, which indicate isotropic magnetic behavior above 5 GPa. Notably, previous studies have suggested that certain magnetic structures, such as the G-type AFM order observed in $MnBi_2Te_4$, may become more stable under high pressure [41,44]. Our findings suggest that G-type AFM order may also become favorable in $MnBi_8Te_{13}$ at comparatively lower pressures (~5 GPa). To conclusively determine the magnetic ground state in this regime, high-pressure neutron diffraction studies will be essential. These investigations would not only clarify the interplay between pressure, exchange interactions, and spin anisotropy but also provide critical insights into the emergence of SC at higher pressures.

**Superconductivity under pressure:**

To further confirm and characterize the SC phase, we measured $R(T)$ under different magnetic fields ($B$ parallel to $c$-axis). A representative dataset recorded at 19.0 GPa is shown in Fig. 6(a). As the magnetic field is increased, the SC transition is progressively suppressed, providing further confirmation that the transition in resistance corresponds to the onset of SC. Here we plot the temperature at which the resistance drops by 10% from the normal-state value under various magnetic fields, and fit the data using both Ginzburg-Landau (G-L) [45] and Werthamer-Helfand-Hohenberg (WHH) [46] models via a universal scaling function [47]. Whereas this criterion may slightly affect the exacted values of $T_c$ and thereby $H_{c2}$, it does not alter the overall trend or the relative fitting quality between the models. A more comprehensive analysis using multiple definitions of $T_c$ could be considered in the future work.

The resulting upper critical field $H_{c2}$ values, determined from fits to the G-L and WHH models, are 15.6 kOe and 11.7 kOe, respectively, at 19 GPa, as shown in Fig. 6(b). Both values are smaller than the Pauli limit of $\mu_0 H_p^{BCS} = 1.84 T_c \approx 88$ kOe. Furthermore, the WHH model, which takes into account both orbital and Pauli limiting effects, provides a better fit to the experimental $H_{c2}(T)$ data compared to the G-L model. This suggests that whereas orbital limiting remains the dominant mechanism for suppressing SC, Pauli paramagnetic effects and potential spin-orbit coupling still play a role in the $H_{c2}(T)$ behavior. The fact that $H_{c2}(0)$ is significantly lower than the Pauli limit indicates that the SC state in $MnBi_8Te_{13}$ is primarily governed by orbital limiting rather than spin-related effects. However, the improved fit of the WHH model over G-L suggests that additional factors, such as electron scattering or multi-band SC, may contribute to the upper critical field evolution with magnetic field.

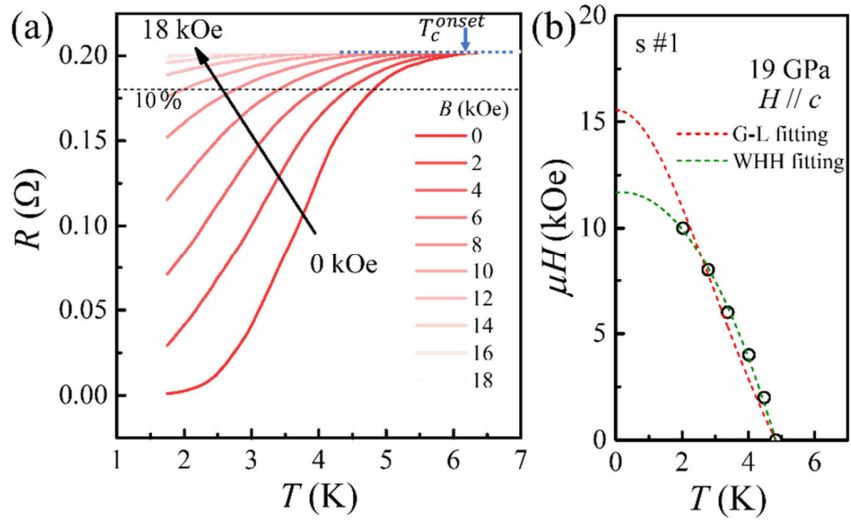

Fig. 6 Superconductivity properties of MnBi$_8$Te$_{13}$ under pressure (a) $R(T)$ for $p$ = 19.0 GPa at different magnetic fields. The criteria of $T_c^{onset}$ defined as is defined as the temperature at which the $R(T)$ curve deviates from the extrapolated linear fit of the normal state resistance; the criteria of $T_c$ used for G-L and WHH model fitting is the temperature at which the resistance drops to 10% of its normal state value. (b) Upper critical fields as a function of temperature determined from $T_c$ at 19 GPa. The red dashed curve is fitted based on the Ginzburg–Landau (G-L) models [45]. The G-L model follows the expression: $H_{c2}(T) = H_{c2}(0)(1-t^2)/(1+t^2)$, where $t = T/T_c$. The green curve is the fit to the Werthamer–Helfand–Hohenberg (WHH) theory [46] using a universal scaling function [47]. (Here we define $T_c$ as the temperature at which the resistance drops to 10% of its normal state value).

**Discussion**

The structural phase diagrams of MnBi$_6$Te$_{10}$ and MnBi$_8$Te$_{13}$, shown in Fig. 1, and the previous results on MnBi$_4$Te$_7$ [23] reveal a remarkable coincidence: all compounds undergo two-step structural transitions at nearly identical critical pressures. The first transition, from rhombohedral ($R$-$3m$) to monoclinic symmetry, emerges near ~10 GPa, followed by a second transition to a high-symmetry cubic structure around ~16 GPa. This consistency-despite differences in layer stacking natures of the different compounds- suggests a shared underlying instability mechanism that is likely in-plane in nature, rather than dominated by the c-axis compression. Interestingly, while all compounds undergo the same structural transitions, the pressure induced superconductivity (SC) was only observed in MnBi$_8$Te$_{13}$

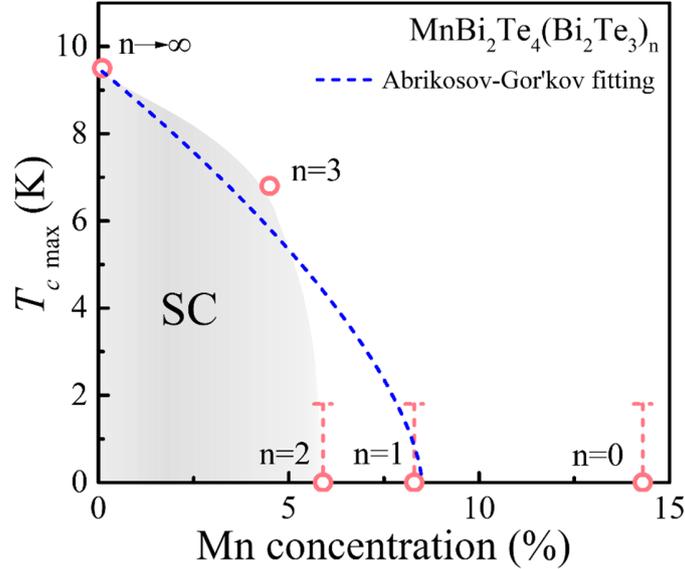

Fig. 7 The relation between the maximum superconducting $T_c$ under high pressure and the unit cell atomic concentration of Mn for MnBi$_2$Te$_4$(Bi$_2$Te$_3$)$_n$ family samples. The blue dashed line is the fitting curve based on the Abrikosov-Gor'kov theory.

Given that all above data suggest a simple structural picture for the Mn-Bi-Te system, the fact that SC under pressure emerges only for n ≥ 3 (Fig. 7), eventually appearing in Bi$_2$Te$_3$ at infinitely high n, while compounds with 0 ≤ n ≤ 2 do not exhibit SC down to 1.8 K within the explored pressure range suggests that SC in the Mn-Bi-Te system is simply linked to the Mn atomic concentration. As n increases, the spacing between Mn layers grows due to the increasing thickness of the Bi$_2$Te$_3$ layers, weakening direct Mn-Mn exchange interactions, suppressing long-range magnetic order and making the system more favorable for SC by decreasing the volume density of Mn moments. So, at higher n, Mn local-like moments become more diluted, reducing pair-breaking effects and allowing SC to emerge, whereas, in lower-n compounds, where Mn layers are more concentrated, strong magnetic correlations persist even under pressure, likely suppressing Cooper pair formation. As such, the Mn-Bi-Te family would be much closer to the RNi$_2$B$_2$C system where the concentration and the local moment on the magnetic, R-ions, govern the rate of suppression of the SC state. [31]

The fitting curve based on Abrikosov-Gor'kov (A-G) theory [48] suggests that $T_c$ in the Mn-Bi-Te family decreases with increasing Mn concentration and that SC could, in principle, exist in lower-n compounds, (n=1 certainly) with a $T_c$ lower than 1.8 K. It should be noted our agreement with AG is qualitative, at best. As was shown for other, similar, magnetic SC, e.g. in the $(Ho_{1-x}Dy_x)Ni_2B_2C$ series, there is a distinct difference between paramagnetic AG scattering and pairbreaking in a magnetically ordered state. [31,48] It does not consider the spin state of the impurity or the exchange interaction between local Mn moments and conduction electrons, both of which can further suppress SC. A-G theory in its simple form does not take into consideration the expected difference between the electronic and phonon structure for different members on the Mn-Bi-Te family. If Mn-induced magnetic order or electronic interactions also play a crucial role in the n = 1 and 2 compounds, the standard A-G model may overestimate the persistence of SC in these cases, explaining the absence of an observed SC transition.

The magnetic and topological character of the $(MnBi_2Te_4)(Bi_2Te_3)_n$ series depends sensitively on the magnetization configurations and symmetry [8]. For small n compounds ($n \leq 2$), the grounds state is an interlayer antiferromagnet (AFM), AFMa [39,40], who combined symmetry $\tau_{1/2}\mathcal{T}$ (time-reversal followed by a half-cell translation a nontrivial bulk $Z_2$ index [49]. For odd layer thin films, both the top and bottom surfaces carry the same half-quantized Hall conductance, yielding an integer Chern number and a quantum anomalous Hall (Chern insulator) phase, whereas for even layer, the surfaces carry opposite half-quantized signs, canceling the net Hall response and realizing a three-dimensional axion insulator with quantized magnetoelectric coupling ($\theta = \pi$). [4,9] In our measurement, we find a ferromagnetic (FM) ground state. As long as the band inversion persists, this magnetic order can still support nontrivial topology in bulk insulators: prior Angle-Resolved Photoemission Spectroscopy (ARPES) studies report gapped, spin-polarized Dirac–like surface bands consistent with an axion insulator. [49] Our sample exhibits typical metallic behaviors throughout the pressure range. We anticipate that Weyl topological states may emerge in these FM phases. [9] With applied pressure, two distinct AFM states emerge. The AFMa state recovers the combined symmetry $\tau_{1/2}\mathcal{T}$ while AFMg2 preserves the combined symmetry $P\mathcal{T}$ ($P$ indicates inversion). These two symmetries protect different 3D topological insulating states: generalized topological insulators and axion insulators. [49-51] Even in the metallic regime, they can all support four-fold degenerate Dirac nodes rather than two-fold degenerate Weyl nodes.

To unravel these topological properties, detailed electronic structures and ARPES measurements are necessary but beyond the scope of this work.

In summary, we have found that the members of the Mn-Bi-Te series have robustly similar critical pressures for pressure induced structural phase transitions and we have uncovered a remarkable three-dome phase diagram in the FM topological insulator candidate $MnBi_8Te_{13}$ under high pressure. Our findings reveal that bulk SC emerges in the sample only after its FM ordering is suppressed and replaced by a second magnetic order, likely AFM. This AFM phase detectable up to approximately 16 GPa, but likely persists to higher pressures with where $T_N \leq T_c$ would be undetectable due to the stabilization of a SC state with a maximum $T_c$ of ~6.8 K. The observed pressure-driven FM-AFM-SC transitions suggest a complex interplay between magnetism, electronic correlations, and SC in this system. Furthermore, the absence of SC in $MnBi_6Te_{10}$ and lower n compounds under similar conditions highlights the critical role of Mn concentration and interlayer coupling in stabilizing the SC phase. Our work establishes $MnBi_8Te_{13}$ as a rare example of a Mn-based SC and provides a new platform for investigating SC in correlated magnetic topological materials. Future studies, including high-pressure neutron diffraction and detailed electronic structure calculations, will be essential to fully understand the nature of the high-pressure magnetic phases and the possible topological characteristics of the SC state in $MnBi_8Te_{13}$.

## Methods:

### Sample growth and characterization

The single crystals of MnBi$_{2n}$Te$_{3n+1}$ (n = 3 and 4) were grown based on self-flux method [10,52,53]. Around 20 g of Mn, Bi, and Te elements with the molar ratio of MnTe:Bi$_2$Te$_3$ as 15:85 were loaded into a 5-ml alumina crucible. The crucible was placed inside a quartz ampoule, and a plug of quartz wool was inserted above the crucible as the filter. The ampoule was then evacuated and sealed under vacuum. The sample ampule was heated up to 900 °C at 200 °C/h, stayed for 5 hours, and then placed inside a separate furnace, which was preheated to 600 °C. Then, it was slowly cooled down to the decanting temperature in 7 days. We then let the ampule dwell at the decanting temperatures for about 7 days before we separated the single crystals from the liquid flux using a centrifuge. Since the decanting temperature is barely above the melting temperature of the flux, the metal ampule holder of the centrifuge was preheated to 500 °C to prevent the flux from solidifying prior to centrifugation. By tuning the decanting temperature, different MnBi$_{2n}$Te$_{3n+1}$ phases can be made dominant. Because the dominant phase is highly sensitive to decanting temperature-a trial-and-error search is recommended to identify the optimal setpoint for each phase on each furnace. As a practical guide: if MnBi$_2$Te$_4$ dominates, lowering the decanting temperature by a few degrees tends to favor MnBi$_4$Te$_7$; lowering it further can shift dominance to MnBi$_6$Te$_{10}$ and then to MnBi$_8$Te$_{13}$. In our setup, MnBi$_6$Te$_{10}$ dominated at 590 °C, whereas MnBi$_8$Te$_{13}$ dominated at 588 °C. To ensure reproducibility despite furnace drift, we calibrated the temperature before each growth and placed the quartz ampoule at the same location in the furnace each time.

### High pressure measurements: resistance and magnetization

The electrical resistance measurements with current applied within the *ab* plane and magnetic field parallel to *c*-axis were performed in a Quantum Design Physical Property Measurement System (PPMS). A standard, linear four-probe method was used for measurement in diamond anvil cell (DAC). We applied 90 kOe and -90 kOe magnetic field to get the precise magnetoresistance (MR) by (*R* (90 kOe, T) + *R* (-90 kOe, T)) / 2. The temperature-dependent resistance measurement up to 40 GPa was performed in a diamond anvil cell (Bjscistar, [54]) that fits into a Quantum Design Physical Property Measurement System (PPMS). 400 μm culet-size standard-cut standard cut-type Ia diamonds were used as anvils. The sample was cleaved, cut and polished into an

100×40×20 μm thin flake and loaded together with a tiny piece of ruby ball into the 250 μm thick, apertured, stainless-steel gasket covered by cubic-BN. The sample chamber was about 150 μm in diameter. Platinum foil was used as the electrodes to connect to the sample. The Nujol mineral oil was used as pressure transmitting medium (PTM), since: 1) fluid medium could still maintain a quasi-hydrostatic pressure environment with small pressure gradient below a liquid/glass transition [55-58]; 2) the use of fluid medium avoids the direct contact between the sample and diamond culet which will lead to a further contribution of uniaxial pressure component. Pressure was determined by $R_1$ line of the ruby fluorescent spectra. [59,60] For the MR measurements with the external magnetic field along different crystallographic orientations, a miniature DAC (the diameter of cell body cylinder is about 12mm) [61] was installed on a lab made device that can manually rotate the direction of the DAC. The rest of the conditions are identical with the above resistance measurements.

The DC magnetization measurements under high pressure were performed in a Quantum Design Magnetic Property Measurement System (MPMS) at temperatures down to 1.8 K. The DAC (easyLab® Mcell Ultra [62]) with a pair of 500-μm-diameter culet-sized diamond anvils was used. The apertured Tungsten gasket with 300-μm-diameter hole was used to lock the pressure. The Nujol mineral oil, same as in the DAC for electrical transport measurement, was used to keep the consistency of pressure environments in two sets of experiments. The applied pressure was measured by the fluorescence line of ruby ball [59,60]. The background signal of the DAC without sample was measured under 3.4 GPa in a 0.2 kOe applied field. Then the DAC was opened and re-closed after loading the sample with a dimension of 150 μm ×150 μm ×20 μm. The exact same measurements as previous background measurements were then performed at various pressures. The magnetization of the sample was analyzed by first performing a point by point subtraction of long-scan response with/without the sample, and then a dipole fitting of the subtracted long-scan response curve [63].

As shown in the phase diagrams of Fig. 1, different sets of electrical resistance data exhibit qualitatively similar pressure-dependent evolution of the ground states, but with slight shifts in the critical pressures. Such variations are likely attributable to differences in the placement of ruby spheres in the pressure cells, which lead to small variations in the local pressure calibration.

**High pressure X-ray diffraction (XRD)**

Room-temperature high-pressure powder XRD (PXRD) measurement was carried out at the GSECARS 13-BM-C station of the Advanced Photon Source (APS), at Argonne National Laboratory (ANL) and 12.2.2 Beamline of Advanced Light Source (ALS), at Lawrence Berkeley National Laboratory, with the wavelengths of x-ray of 0.4306 Å and 0.4956 Å, respectively. The crystalline sample was grounded into powder and loaded into a wide opening SSDAC-70 DAC. Diamond anvils with 300 μm diameter culet and rhenium gasket were used to form the sample chamber. Neon gas was loaded as the PTM. A ruby sphere (~30 μm in diameter) was loaded next to the sample to serve as the pressure marker. Pressures were determined *in-situ* using the ruby fluorescence method [59,60]. The two-dimensional diffraction images were collected using the Pilatus 1M area detector and were integrated using the DIOPTAS software [64]. Rietveld and LeBail refinements were performed in GSAS-II [65].

It is noteworthy that the use of different pressure-transmitting media, powder PXRD employed Neon gas, whereas both electrical resistance and magnetization measurements used Nujol mineral oil, may lead to discrepancies in the determination of the boundaries of different phases. Such differences in hydrostatic conditions can influence the accuracy and comparability of the phase diagrams plotted from different experimental probes.

**DFT calculation**

Density functional theory [66,67] (DFT) calculations have been performed with van der Waals interaction [68] of D3, a plane-wave basis set and projected augmented wave method [69] as implemented in VASP [70,71]. A Hubbard-like U [72] value of 4.0 eV is used to account for the half-filled strongly localized Mn 3$d$ orbitals. To compare the phase stability of different magnetic configurations of MnBi$_8$Te$_{13}$, a 176-atom supercell has been constructed to accommodate the anti-ferromagnetic G-type double-row strip (AFMg2) besides A-type AFM (AFMa) and ferromagnetic (FM) configurations. DFT total energies are calculated for all three magnetic configurations in the supercell with a Monkhorst-Pack [73] (12 × 3 × 1) $k$-point mesh including the Γ point and a Gaussian smearing of 0.05 eV for a kinetic energy cutoff of 337.3 eV. The shape of the supercell and ionic positions were relaxed until the absolute value of force on each atom is less than 0.002 eV/Å for the volumes scaled around the equilibrium. Then these results were fit to Birch-Murnaghan [74] equation of state to evaluate pressure.

**Acknowledgements**

Work at Ames National Laboratory is supported by the U.S. Department of Energy (DOE), Basic Energy Sciences, Material Science and Engineering Division under Contract No. DE-AC02-07CH11358. S.H. was supported in part by the Ames National Laboratory's Laboratory Directed Research and Development (LDRD) program. Work at UCLA was supported by the U.S. Department of Energy (DOE), Office of Science, Office of Basic Energy Sciences under Award No. DE-SC0021117. W.B. acknowledges support from NSF CAREER Award No. DMR-2045760. F.X. acknowledges the support by the National Science Foundation under Grant No. OIA-2229498. Portions of this work were performed at GeoSoilEnviroCARS (The University of Chicago, Sector 13), Advanced Photon Source, Argonne National Laboratory. GeoSoilEnviroCARS is supported by the National Science Foundation– Earth Sciences via SEES: Synchrotron Earth and Environmental Science (EAR–2223273). Portions of high pressure XRD work were performed at Beamline 12.2.2, Advanced Light Source, Lawrence Berkeley National Laboratory, a U.S. DOE Office of Science User Facility under contract no. DE-AC02-05CH11231. B.K. acknowledges SEES supported by the National Science Foundation Division of Earth Sciences (EAR) SEES: Synchrotron Earth and Environmental Science (EAR –2223273). Z.L. and T.J.S. were partially supported by the Center for Advancement of Topological Semimetals (CATS), an Energy Frontier Research Center funded by the U.S. Department of Energy Office of Science, Office of Basic Energy Sciences, through Ames National Laboratory.

Supplementary Information for

# Pressure-Induced Changes in Structure, Magnetic Order and Development of Superconductivity in the Ferromagnetic Topological Insulator MnBi$_8$Te$_{13}$

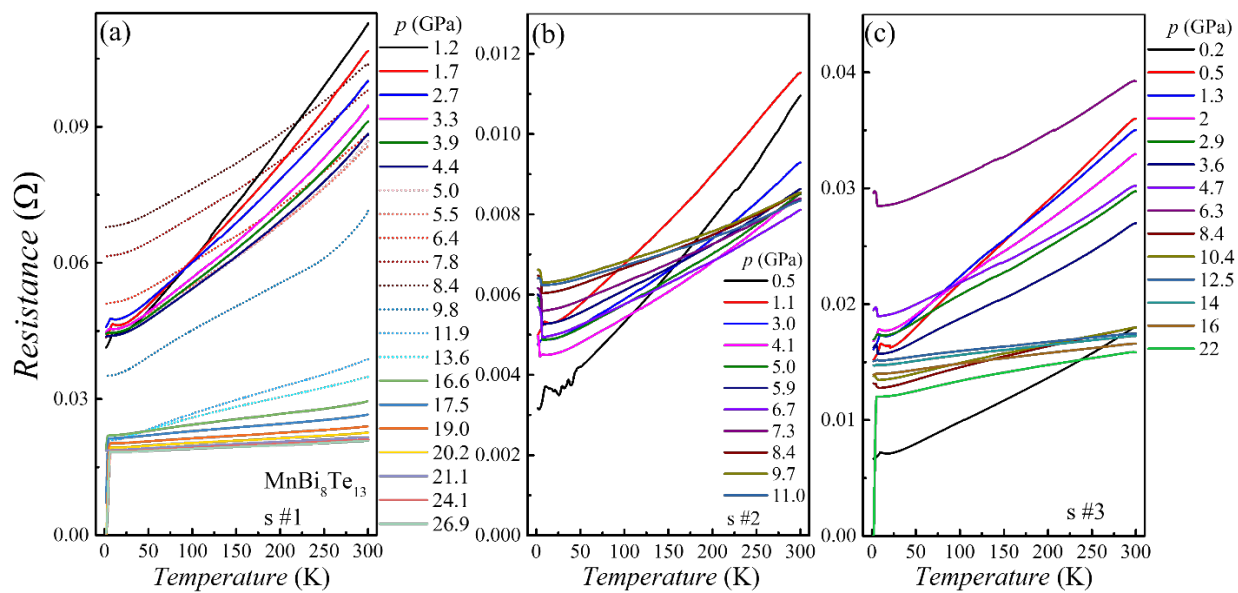

Fig. S1 Resistance as a function of temperature at various pressures on MnBi$_8$Te$_{13}$ single crystals (a) s #1, (b) s#2, and (c) s #3.

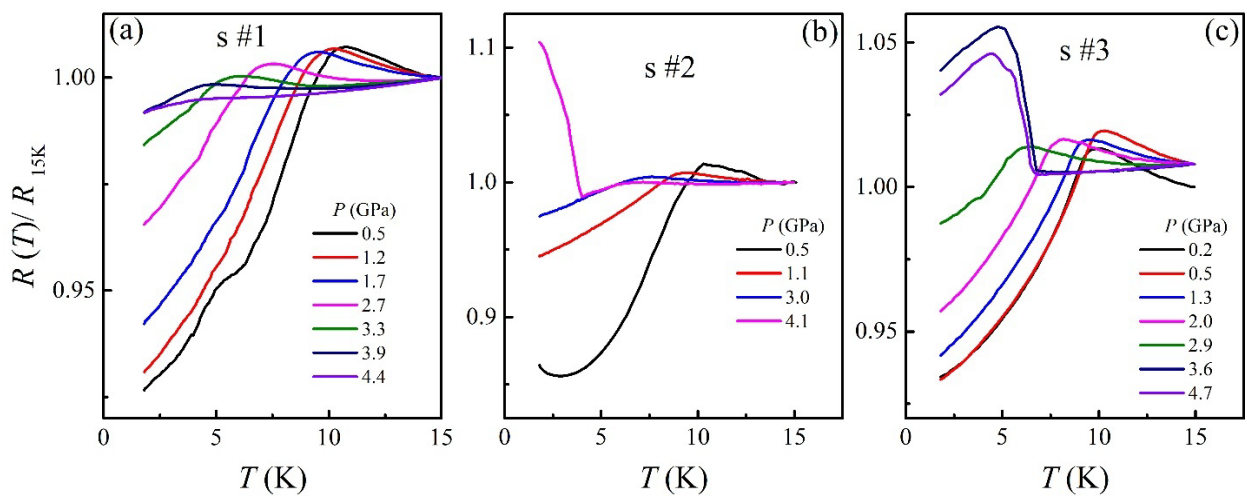

Fig. S2 $R(T)/R_{15K}$ as function of temperature under pressure in the low pressure range (< 5 GPa) for 3 MnBi$_8$Te$_{13}$ samples.

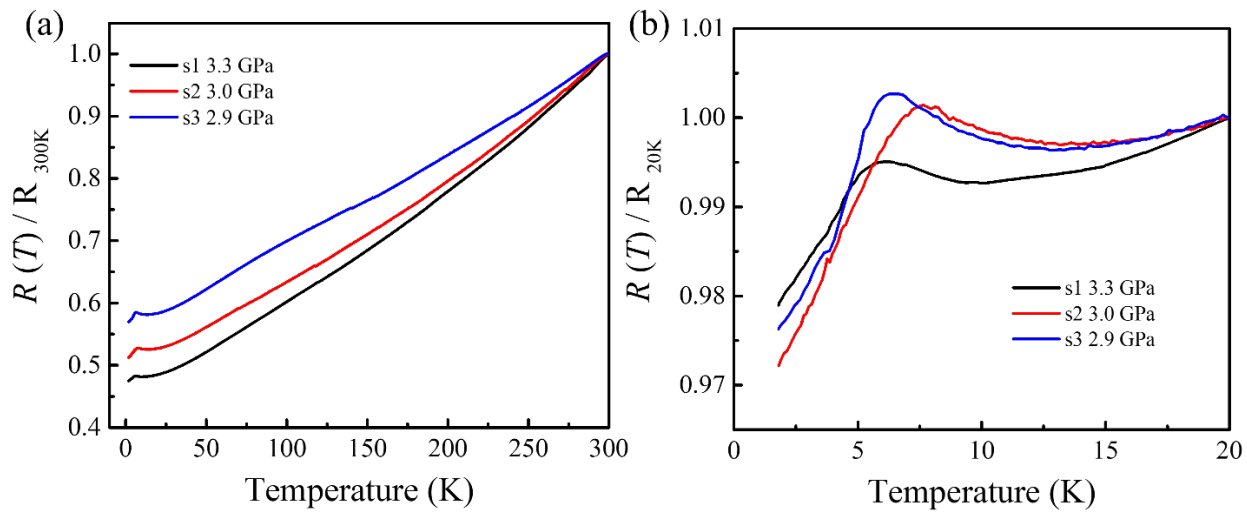

Fig. S3 Resistance vs. temperature plots for three different samples (S#1, S#2, and S#3) measured at ~3 GPa. (a) Full temperature range and (b) expanded view below 20 K for low-temperature comparison.

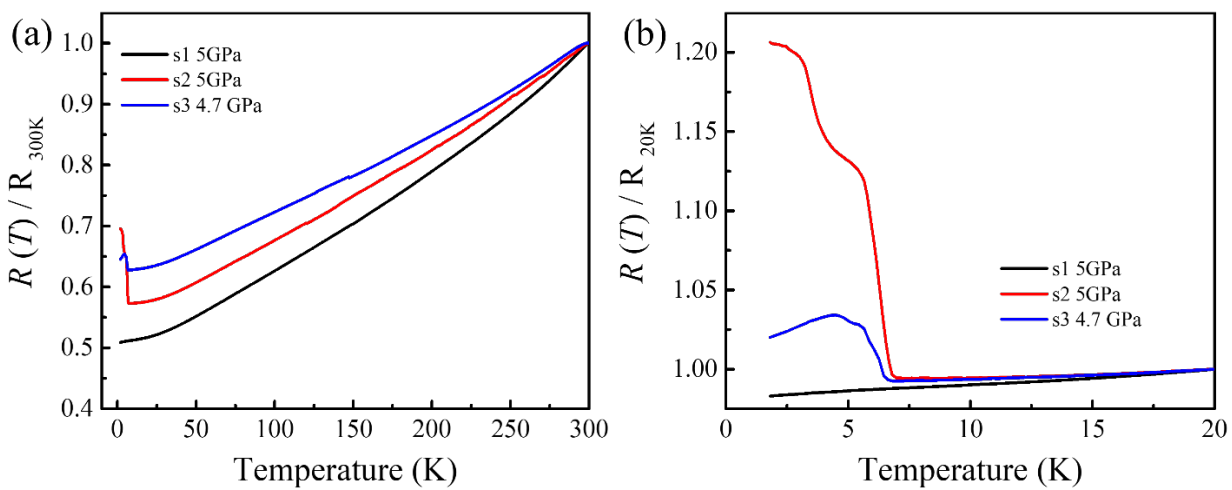

Fig. S4 Resistance vs. temperature plots for three different samples (S#1, S#2, and S#3) measured at ~5 GPa. (a) Full temperature range and (b) expanded view below 20 K for low-temperature comparison.

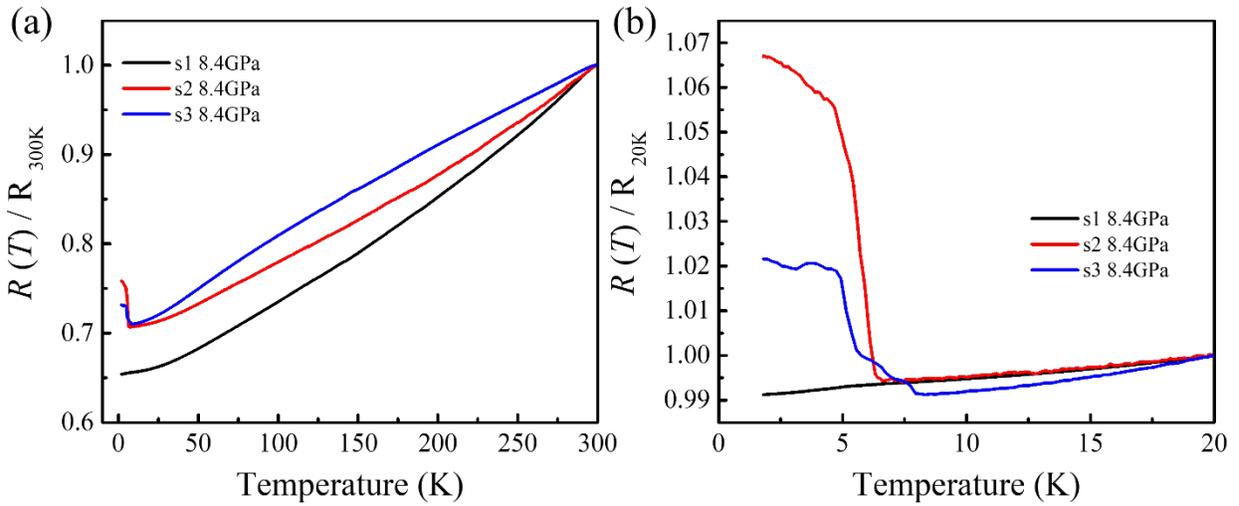

Fig. S5 Resistance vs. temperature plots for three different samples (S#1, S#2, and S#3) measured at ~8.4 GPa. (a) Full temperature range and (b) expanded view below 20 K for low-temperature comparison.

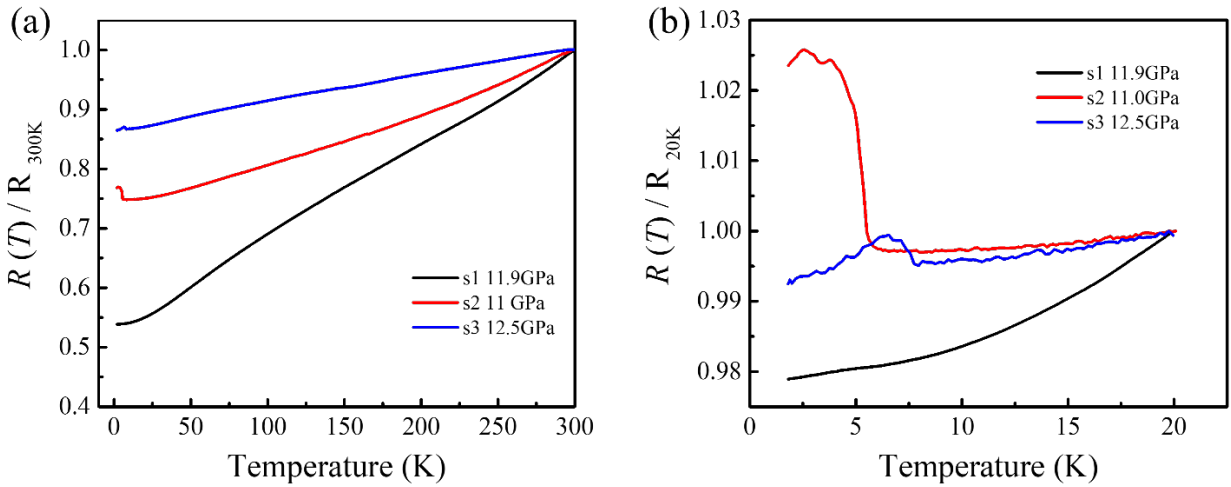

Fig. S6 Resistance vs. temperature plots for three different samples (S#1, S#2, and S#3) measured at ~12 GPa. (a) Full temperature range and (b) expanded view below 20 K for low-temperature comparison.

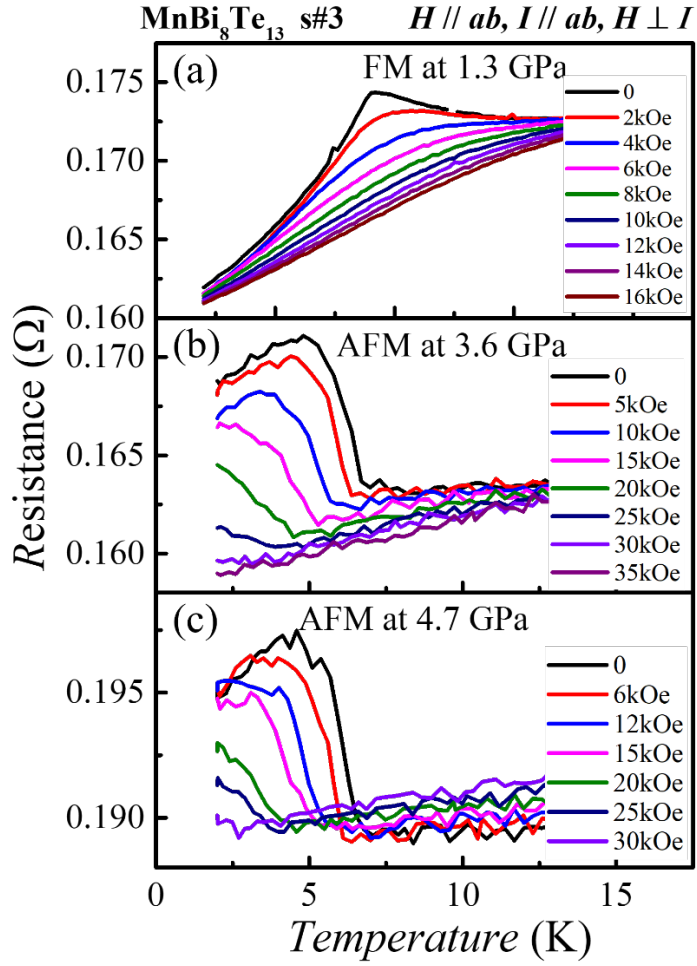

Fig. S7 (a), (b) and (c) temperature dependence of the resistance at various magnetic fields, at 1.3 and 3.6 GPa, and 4.7 GPa, respectively, on sample #3. The magnetic field is along *c* direction. The current is parallel to *ab* plane.

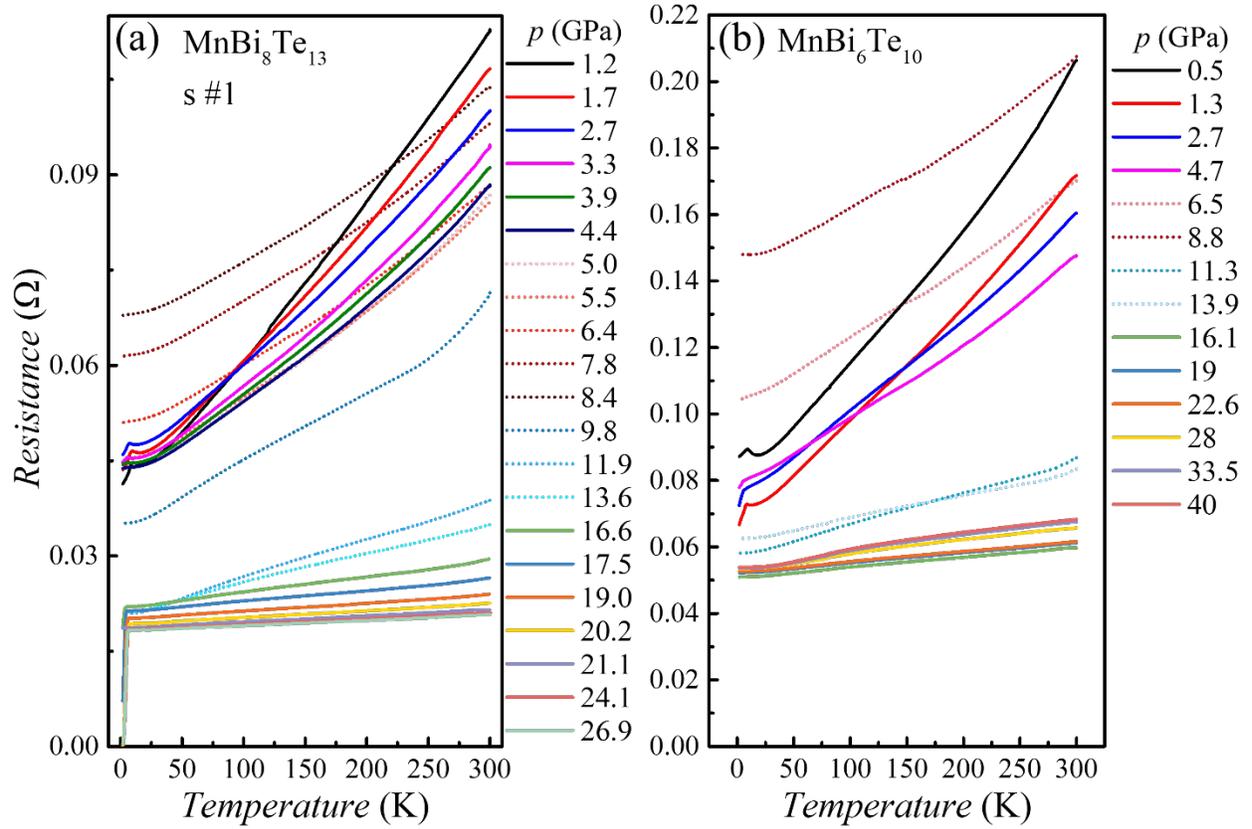

Fig. S8 Electrical resistance as a function of temperature at various pressures on (a) MnBi$_8$Te$_{13}$ (s #1), and (b) MnBi$_6$Te$_{10}$, single crystals.

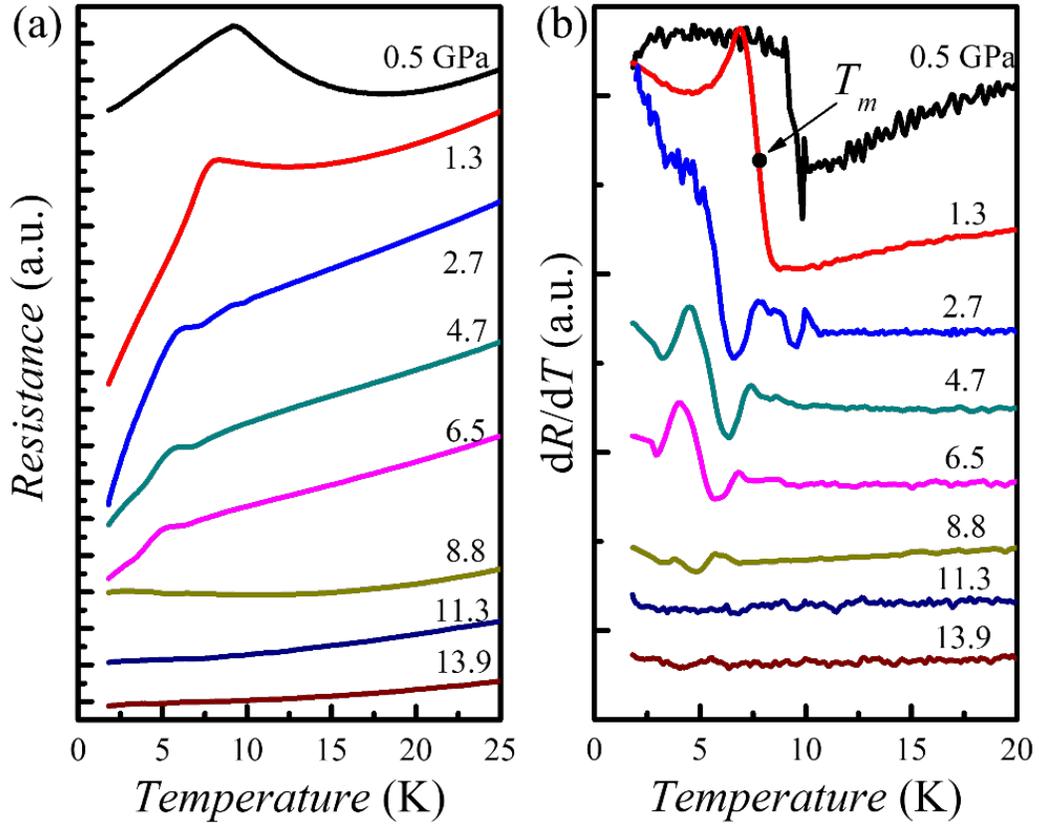

Fig. S9 (a) Electrical resistance as a function of temperature at various pressures in the low-temperature regime, on MnBi$_6$Te$_{10}$ single crystal. (b) Corresponding temperature derivatives (d$R$/d$T$) under different pressures, indicating the magnetic transition. The transition temperature is defined by the midpoint of the d$R$/d$T$ shoulder, indicated by solid black spot.

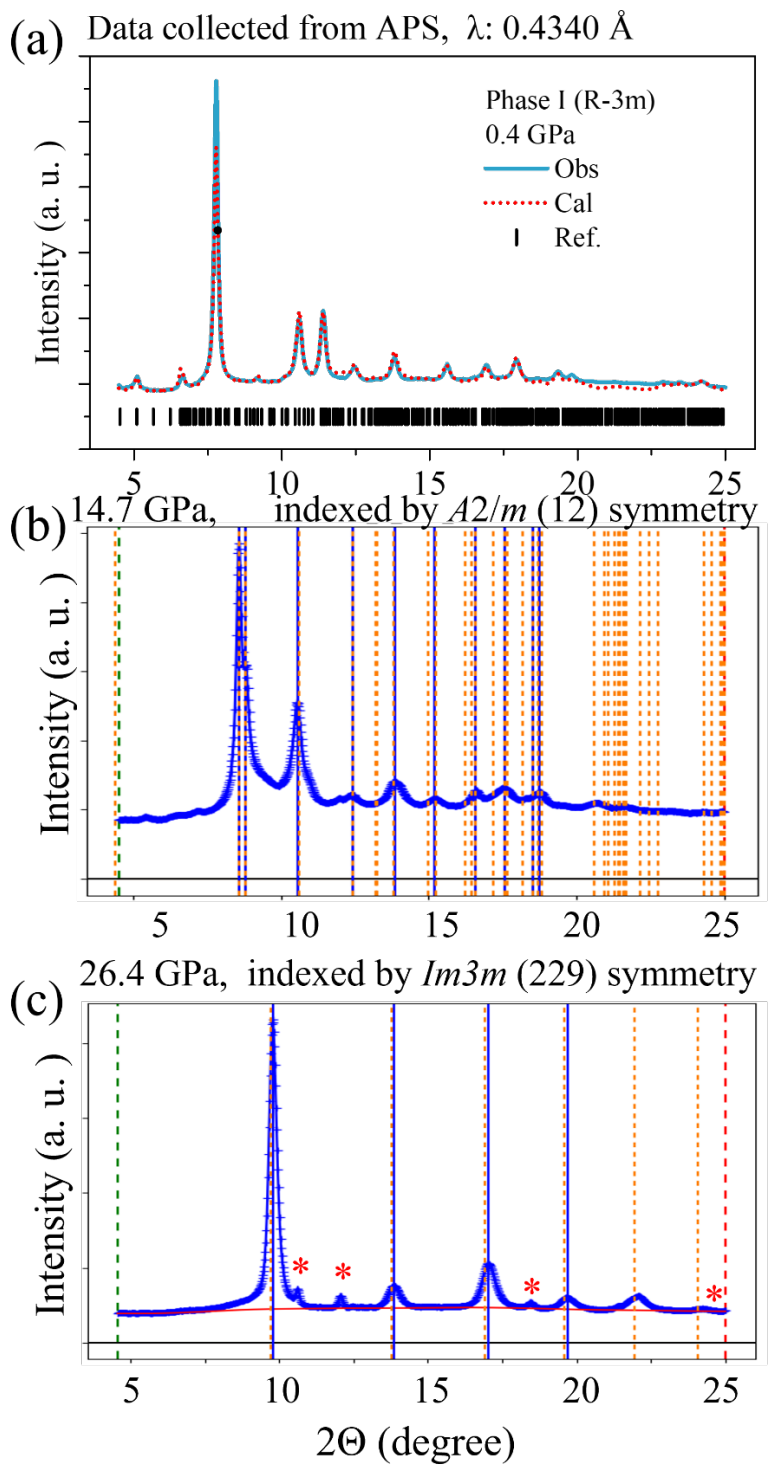

Fig. S10 The detailed X-ray diffraction data on 3 different phases for MnBi$_8$Te$_{13}$. (a) Rietveld refinement of the PXRD pattern at 0.4 GPa. (b) The indexing of X-ray diffraction spectrum at 14.7 GPa with $A2/m$ group symmetry. (c) The indexing of X-ray diffraction spectrum at 26.4 GPa with $Im3m$ group symmetry. The red star in (c) represent the clear extra peaks in the spectrum of rhenium.